\documentclass[preprint2]{aastex}

\begin{document}


\title{Classical and Quantum Reissner-Nordstr\"{o}m Black Hole Thermodynamics and first order Phase Transition}


\author{Hossein Ghaffarnejad \altaffilmark{1}} \affil{Faculty of
Physics, Semnan University, Semnan, Zip Code: 35131-19111, IRAN}
\altaffiltext{1}{hghafarnejad@yahoo.com\\
hghafarnejad@profs.semnan.ac.ir
}



\begin{abstract}
First we consider classical Reissner-Nordstr\"{o}m black hole
(CRNBH) metric which is obtained by solving  Einstein-Maxwell
metric equation for a point electric charge $e$ inside of a
spherical static body with mass $M$. It has 2 interior and
exterior horizons.  Using Bekenestein-Hawking entropy theorem we
calculate interior and exterior entropy, temperature, Gibbs free
energy and
 heat capacity at constant electric charge.
 We calculate first derivative of the Gibbs free energy with respect to temperature which become a singular function having a singularity
 at critical point $M_c=\frac{2|e|}{\sqrt{3}}$ with corresponding temperature $T_c=\frac{1}{24\pi \sqrt{3}|e|}.$
  Hence we clime first order phase transition is happened there.
Temperature same as Gibbs free energy takes absolutely positive
(negative) values on the exterior (interior) horizon. The Gibbs
free energy takes two different positive values synchronously for
$0<T<T_c$ but not for negative values which means the system is
made from two subsystem. For negative temperatures entropy reaches
to zero value at $T\to-\infty$ and so takes Bose-Einstein
condensation single state. Entropy increases monotonically in case
$0<T<T_c$. Regarding results of the work presented at Ref.
\citep{Bob01} we calculate again the mentioned thermodynamical
variables for remnant stable final state of evaporating quantum
Reissner-Nordstr\"{o}m black hole (QRNBH) and obtained results
same as one in case of the CRNBH. Finally, we solve mass loss
equation of QRNBH against advance Eddington-Finkelstein time
coordinate and derive luminosity function. We obtain switching off
of QRNBH evaporation before than the mass completely vanishes. It
reaches to a could Lukewarm type of RN black hole which its final
remnant mass is $m_{final}=|e|$ in geometrical units. Its
temperature and luminosity vanish but not in Schwarzschild case of
evaporation. Our calculations can be takes some acceptable
statements about information loss paradox (ILP).
\end{abstract}





\section{Introduction}
 Since the seminal work of Hawking  \citep{Haw74} and Bekenstein \citep{Bek73}, we have
 understood that black holes behave as thermal objects containing  characteristics such as
 temperature, entropy and et cetera.
  Hawking radiation has
 not yet been directly observed, of course; a typical stellar mass
 black hole has a Hawking temperature of well under a micro-Kelvin,
 far lower than that of the cosmic microwave background temperature $\approx2.7 K$. However the
 thermal properties of black holes are studied in the literature and there is
 well understood that they have temperature
 \begin{equation} k T_{Hawking}=\frac{\hbar \kappa}{2\pi}\end{equation}
and entropy \begin{equation}
S_{BH}=\frac{A_{horizon}}{4G\hbar}.\end{equation} In the above
equations $`A_{horizon}`$ is event horizon surface area, $`\kappa`$
is corresponding surface gravity which is constant over the event
horizon, $G$ is Newton`s coupling constant and $\hbar$ is Plank
coupling constant. One can obtain value of the surface gravity by
applying second law of the black hole thermodynamics:
\begin{equation} \delta M=\frac{\kappa}{8\pi G}\delta A+\Omega_H\delta
J+\Phi_H\delta e\end{equation} where $`\Omega_H`$is angular
velocity and $`\Phi_H`$ is electric  potential given on the
horizon. Also `$e$` and `$M$` is electric charge and mass of the
black hole respectively. Surface area of the event horizon of a
black hole never decreases such that \begin{equation}\delta
A\geq0.\end{equation} In a typical thermodynamic system,
temperature is a measure of average energy of microscopic
constituents and entropy counts the number of microstates. However
one can see that the Bekenstein-Hawking entropy depends on both
Plank`s and Newton`s constant and so obtain a good statement:
Statistical mechanic description of black hole thermodynamics
might be tell us something profound about quantum gravity
\citep{Val98,Pap09}. A compendious  review about the black hole
thermodynamics and its relation with other topics as quantum
gravity, generalization of thermodynamic laws, statistical
properties and information loss paradox is given in ref.
\citep{Car15} (see also references therein).\\
This encourages us to study some thermodynamical aspects about the
CRNBH metric. Here we calculate its entropy $A(M,e)$, temperature
$T(M,e)$, Gibbs free energy $G(M,e)$ and heat capacity $C_e(M,e)$
at constant charge $e$ on the interior (Cauchy) and exterior
horizons. We obtain corresponding temperature takes negative
(positive) values for interior (exterior) horizon of the CRNBH
metric (see figures 1 and 2). For negative temperatures on the
interior horizon the Gibbs free energy takes negative values but
for positive temperatures will be have positive values (see figure
3). Entropy and so the black hole microstates degrees of freedom
decreases by increasing initial mass $M$ (see dot line at figures
1 and 2). This happened for negative temperatures (see dash line
at figures 1 and 2). Thus we can result the entropy reaches to
zero value when negative temperatures approach to negative
infinity and so the black hole can be takes Bose-Einstein
condensation state with zero momentum and minimum energy (see
figure 4). We obtain corresponding heat capacity calculated on
exterior horizon which exhibits with discontinuity at critical
point $M_c=\frac{2|e|}{\sqrt{3}}.$ This means that phase
transition is happened there (see solid line at figure 1).
\\
Phase transition phenomena and Bose-Einstein condensation state
are well known properties in statistical systems. For instance a
$\Lambda$-transition type is happened usually for isotope $He^4$
of liquid helium at low critical temperatures as $2.17 K.$ The
isotope $He^3$ has nuclear spin $\frac{1}{2}$ and so obeys
Fermi-Dirac statistics, while the isotope $He^4$ with nuclear spin
$0$ obeys Bose-Einstein statistics. At very low temperatures where
quantum effects become important, $He^3$ and $He^4$ as quantum
fluids will have identical chemical properties but with different
masses or energies. Two isotopes $He^3$ and $He^4$ exhibit very
different behavior due to the difference in their statistics.
Liquid $He^4$ which is a boson liquid exhibits at rather
straightforward transition to a superfluid state at 2.19 Kelvin.
This can be understood as a (Bose-Einstein) condensation of
particles into a single quantum state with zero momentum. Liquid
$He^3$ also undergoes at transition to a superfluid state, but at
a much lower temperature $(2.7\times10^{-3})$ Kelvin. The
mechanism for a latter isotope is different from the liquid $He^4$
where quasi-particles form bound pairs with spin $s=1$ and
relative angular momentum $l=1.$ One can see experimental diagram
of $\Lambda$-transition for liquid $He^4$ system at figure 7.5 of
ref. \citep{Pat72} or figure 4.22 of ref. \citep{Rei80} (see also
page 153 in ref. \citep{Hua01} ).\\
 According to the latter
phenomena about the liquid helium we see that RHS of figure 3 in
this article shows that matter content of the CRNBH namely $M$
takes two different Gibbs free energy at temperatures $0<T<T_c.$
This means
 the CRNBH will be has two
different phases (two subsystem) at particular temperature given
by $0<T<T_c.$ These 2 phases reach to a single state at critical
temperature $T_c=\frac{1}{24\sqrt{3}|e|}$ described by same Gibbs
free energy $G_c=\frac{|e|}{2\sqrt{3}}$. LHS of the figure 3 shows
that the CRNBH takes Bose-Einstein condensation state when
temperature approaches to negative infinite values $T\to-\infty,$
where entropy takes zero value and so microstates degrees of
freedom vanishes. This happened on the interior horizon of the
CRNBH metric.\\ Gibbs free energy is the chemical potential that
is minimized $\Delta G=0$ when a system reaches equilibrium at
constant pressure and temperature. In other words its derivative
with respect to the reaction coordinate of the system vanishes at
the equilibrium point. A reaction with a negative (positive) Gibbs
free energy will (not) proceed spontaneously. In other words when
$\Delta G<0 (\Delta G>0)$, the system will be has spontaneous
(non-spontaneous) reactions with constant pressure and
temperature.\\
 Negative temperatures have physical
meaning and is possible for ordinary statistical systems if there
exists an upper limit for the energy of the given system. Ordinary
systems such as freely moving particle or a harmonic oscillator
possess usually kinetic energy of motion which is obviously
unbounded and so can not be suitable candidates for systems with
negative temperature. In these ordinary translational and
vibrational degrees of freedom (and so entropy) of a body
increases without limit as the energy increases. Two-level spin
systems for instance magnetic dipoles in presence of external
magnetic field are suitable candidates where the system exhibits
with negative temperature when population of dipoles with higher
energy state is more than that in the lower energy state (see for
instance section 3.9, ref. \citep{Pat72}). For these systems
variations of entropy with respect to their energy takes negative
sign $\Delta A<0$ (for instance see dot line in figure 2 given in
this work). \\ In summary, our study about thermodynamical aspect
of the CRNBH metric predicts that it is formed from two subsystem:
(a) Matter content inside of interior horizon $0<r<r_-$ with
negative temperatures, reaching to Bose-Einstein condensation
state and (b) matter content inside between interior and exterior
horizon $r_-<r<r_+$ with positive temperatures exhibiting to a
first order phase transition.\\
At second part of the paper we extend aim of the work to a QRNBH
metric by using results of the work presented by \citep{Bob01}. We
calculate again its entropy, temperature, Gibbs free energy and
heat capacity at constant electric charge. Bobo et al solved
time-independent backreaction metric equation of quantum perturbed
RN black hole in the presence of massless, charge-less quantum
scalar field. They obtained metric of final state as remnant
stable static black hole which its horizon radiuses are greater
than the classical horizon radiuses $r^Q_{\pm}>r_{\pm}.$
\\
Our mathematical calculations for QRNBH in static regime of the
backreaction metric solution predict results same one which
obtained in cases of CRNBH metric(see figures 5, 6, 7, 8 and
compare with figures 1,2,3,4). Also we calculate luminosity of evaporating QRNBH and obtained
a switching off for it before that its mass function disappear completely.
Also we give some suitable statement about ILP.
  Organization of the paper is given as follows.\\
In section 2 we call CRNBH metric as static solution of
Einstein-Maxwell equation. Also we discuss its stress tensor which
treats as an anisotropic hydrostatic fluid with negative
barotropic and anisotropy indexes. Also we call its internal and
external horizon radiuses. In section 3 we calculate entropy,
positive temperatures , Gibbs free energy and heat capacity on its
exterior horizon. In section 4 we calculate entropy, negative
temperature, Gibbs free energy and heat capacity on its interior
horizon. Also we discuss about its Bose Einstein condensation
state. In section 5 we call QRNBH dressed interior and exterior
quantum perturbed interior and exterior horizons by using results
of the work Bobo et al \citep{Bob01}. We obtain time dependent
mass function by solving mass loss equation. Also we calculate
luminosity of the QRNBH and give some acceptable physical
statements about ILP. Section 6 denotes to concluding remark.
\section{Classical Reissner Nordstr\"{o}m Black Hole}

CRNBH is  metric solution of Einstein-Maxwell equation
$G_{\mu\nu}=8\pi T^{EM}_{\mu\nu}$ where
\begin{equation}
T_{\mu\nu}^{EM}=-F_{\mu\alpha}F^{\alpha}_{\nu}+\frac{1}{4}g_{\mu\nu}(F_{\alpha\beta}F^{\alpha\beta})\end{equation}
is electromagnetic fields stress tensor of antisymmetric
electromagnetic tensor field $F_{\mu\nu}=\partial_{\mu}
A_{\nu}-\partial_{\nu}A_{\mu}.$ The above stress tensor is written
with metric signature (-,+,+,+) by setting $c=G=\hbar=1.$ It is
trace free as $g^{\mu\nu}T^{EM}_{\mu\nu}=0$ because the
electromagnetic fields propagate at the invariant speed $c$
evaluated in all reference frames by all observers. The invariance
of speed of light is one of postulates of special (and also
general) relativity theory. In various alternative gravity
theories` variable speed of light` (VSL) is accepted as hypothesis
because velocity of the light rays take slow down when traveling
through a medium. VSL should not be confused with faster than
light theories. Notable VSL attempts have been done by Einstein
itself  \citep{Ein07} and other researchers as Robert Dicke
\citep{Dic57} (see also \citep{Cla99}).\\ In case of unmoving
point particle with electric charge $e$ and negligible mass
$\mathfrak{m}_e<<M,$ located inside of spherical body with mass
$M,$ one can obtain nonzero components of the electromagnetic
fields as $F_{tr}=-F_{rt}=E(r)=\frac{e}{r^2}$  in units
$\frac{1}{4\pi \epsilon_0}=1.$ In the latter case the
electromagnetic field stress tensor (5) takes a simple form as
\begin{equation}
{T^{(EM)}}^{\mu}_{\nu}=\frac{1}{8\pi}\left(\frac{e}{r^2}\right)^2diag\left(%
\begin{array}{cccc}
  1 & 1 & -1 & -1 \\
\end{array}%
\right).
\end{equation}
If we assume that the above electric field stress tensor treats
same as anisotropic hydrostatic perfect fluid, then corresponding
effective density will be
\begin{equation}
\rho_{eff}=T^t_t=\frac{e^4}{8\pi r^4},
\end{equation} effective radial pressure become
 \begin{equation}
p^r_{eff}=T^r_r=\frac{e^4}{8\pi r^4}
\end{equation}
and effective transverse (tangential) pressure is
\begin{equation}
p^t_{eff}=T^{\theta}_{\theta}=T^{\varphi}_{\varphi}=-\frac{e^4}{8\pi
r^4}.\end{equation} Corresponding hydrostatic pressure $p$ is
defined by \begin{equation} p=\frac{p^r+2p^t}{3}=-\frac{e^4}{24\pi
r^4}\end{equation} and anisotropic stress tensor  is defined by
\begin{equation}\Pi^{\mu}_{\nu}=(p^r-p^t)diag\bigg\{0,\frac{2}{3}, -\frac{1}{3},-\frac{1}{3}\bigg\}$$$$=
\frac{e^4}{4\pi r^4}diag\bigg\{0,\frac{2}{3},
-\frac{1}{3},-\frac{1}{3}\bigg\}.
\end{equation} Inserting (7), (8), (9) and (10), one can obtain corresponding
barotropic and anisotropy indexes respectively as
\begin{equation}\gamma=\frac{p_{eff}}{\rho_{eff}}=-\frac{1}{3}\end{equation}
 and \begin{equation}
\Delta=\frac{p_{eff}^t-p_{eff}^r}{\rho}=-2.\end{equation} The
barotropic index (12) has negative value and so negative pressure
of electric field of point particle produces anti-gravity
(repulsive force), relative to attractive force of the mass $M$ in
the CRNBH metric
\begin{equation}
ds^2=-\left(1-\frac{2M}{r}+\frac{e^2}{r^2}\right)dt^2+\frac{dr^2}{\left(1-\frac{2M}{r}
+\frac{e^2}{r^2}\right)}$$$$+r^2(d\theta^2+\sin^2\theta
d\varphi^2).\end{equation} The above metric solution is obtained
by inserting (6) and solving the Einstein-Maxwell equation
$G_{\mu\nu}=8\pi T^{EM}_{\mu\nu}$ for a spherically symmetric
static line element
\begin{equation}ds^2=-A(r)dt^2+B(r)dr^2+r^2(d\theta^2+\sin^2\theta d\varphi^2).\end{equation} The metric
solution (14) is asymptotically flat and called as metric of
CRNBH. Exterior and interior  event horizons is obtained by
solving $g_{tt}(r)=0$ as $r_+=M+\sqrt{M^2-e^2}$ and
$r_-=M-\sqrt{M^2-e^2}$ respectively but its apparent horizon is
obtained from equation $g^{\mu\nu}\partial_\mu r\partial_\nu r=0$
as $ r_{\pm}=M\pm\sqrt{M^2-e^2}.$ Its interior (exterior) event
horizon coincide with interior (exterior) apparent horizon.
Interior horizon is called usually as `Cauchy` horizon and can not
be seen via observers located outside of exterior horizon. All
horizons exist for $0\leq \frac{|e|}{M}\leq1$ and for
$\frac{|e|}{M}>1$ the horizons are disappeared and so the metric
solution (14) takes a naked singularity form. With particular
choice $|e|=M$ (Lukewarm) we have $r_-=r_+=M.$ With $|e|=0$ the
metric solution (14) leads to Schwarzschild space time where $r_+=2M,~r_-=0$ \citep{Wal84,Ghaet13}. \\
It is well known, negative values for barotropic index of an
arbitrary fluid is usually related to dark matter and dark energy
which can be support carefully acceleration of our expanding
universe. Also a true cosmological constant $\Lambda$ may be
responsible for the data as $\gamma_{\Lambda}=-1$ but it is
possible that a dynamical mechanism is at work. Dark energy has
several candidates as quintessence and non-canonical (negative
kinetic energy) k-essence scalar fields. Experimental tests
determine $-1.38<\gamma_{DE}<-0.82$ for the barotropic index of
dark energy by using WMAP data and CMB experiments \citep{Mel et
al03}. On the other hand strong evidence, from a large number of
independent observations indicates that dark matter is composed by
yet un-known weakly interacting elementary particles. Since these
particles are required to have small random velocities at early
times, they are called cold dark matter (CDM) with barotropic
index $-1<\gamma_{CDM}<0$ \citep{Ser et al11}. The equation (12)
accords the dark matter barotropic index and so one can obtain a
correspondence between electric field stress tensor of point
particle (6) and unknown CDM stress tensor. However, physical
effects of the electric charge stress tensor (6) cases to break
Schwarzschild black hole horizon radius $r_{Sch}=2M$ to two
different horizon radiuses called as interior horizon
$r_-=M-\sqrt{M^2-e^2}$ and exterior horizon $r_+=M+\sqrt{M^2-e^2}$
for CRNBH which both are less than the Schwarzschild one.
\\
In the following we will restrict ourselves to a typical situation
$e^2<M^2$ for which the horizons are not destructed and so there
is not naked singularity. Physically the condition $e^2<M^2$ means
the black hole has a remnant case and so may have implications
about ILP. We will discuss about ILP at last section of the paper.
We calculate now entropy temperature, Gibbs free energy and heat
capacity of CRNBH metric (14) at constant electric charge for both
interior and exterior event horizons.
\section{Entropy on exterior
horizon} The CRNBH is static  with no angular momentum
$\Omega_H=0$ and so with no electric current density $J=0$. Thus
it can not contribute to electromagnetic interactions and its
electric charge will be invariant with no electromagnetic
radiation but can be contribute with gravitational and
electrostatic self-interactions only. With these boundary
conditions one can rewrite (3) for the CRNBH as
\begin{equation} \delta M=T\delta A+\Phi\delta e\end{equation} where $A, T, M$ and $\Phi$ are entropy, temperature, mass and electric potential
 respectively defined on horizons hyper-surface. First we calculate mentioned
 thermodynamical functions on the exterior horizon of CRNBH metric as follows. \subsection{Positive temperatures}
Inserting exterior horizon radius $r_+=M+\sqrt{M^2-e^2},$ surface
area defined by  $A_+=4\pi r_+^2$ become
\begin{equation}A_+(M,e)=4\pi (M+\sqrt{M^2-e^2})^2.\end{equation}
According to the Bekenstein-Hawking entropy theorem the
 equation
(17) describes entropy of exterior horizon counter part of the
CRNBH. Using (16) and varying (17) with respect to $M$ and $e$,
one can obtain corresponding temperature $T_+$ and electric
potential $\Phi_+$ respectively as
\begin{equation}
T_+(M,e)=\frac{\sqrt{M^2-e^2}}{8\pi
\left(M+\sqrt{M^2-e^2}\right)^2}\end{equation} and
\begin{equation}
\Phi_+(M,e)=\frac{e}{M+\sqrt{M^2-e^2}}\end{equation} which become
real functions only for $M^2\geq e^2.$ So the CRNBH should be has
remnant case with mass lower limit as $M_{min}=|e|$. Heat capacity
of exterior horizon at constant electric charge $e$ is obtained as
\begin{equation} C^+_e(M,e)=8\pi
\bigg[\frac{(M+\sqrt{M^2-e^2})^2\sqrt{M^2-e^2}}{M-2\sqrt{M^2-e^2}}\bigg]\end{equation}
by using the definition
\begin{equation}
C^+_e(M,e)=T_+(M,e)\left(\frac{\partial A_+(M,e)}{\partial
T_+(M,e)}\right)_e$$$$=T_+(M,e)\frac{(\frac{\partial
A_+(M,e)}{\partial M})_e}{(\frac{\partial T_+(M,e)}{\partial
M})_e}.\end{equation} Diagram of the solutions (17), (18) and (20)
are plotted against $M$ for particular electric charges $e=\pm1$
in figure 1. Diagram of the heat capacity (20) exhibits with a
singularity (see solid line in figure 1) at critical mass
$M_c=\frac{2|e|}{\sqrt{3}}\simeq1.1547|e|$ and its sign is changed
from positive to negative values \citep{Gha13}. Temperature and
entropy do not exhibit with singularity and take positive values
always for all $M$ (see dash and dot lines respectively in figure
1). Entropy diagram satisfies (4).
\subsection{Phase transition}
First (second) order phase transitions is happened in ordinary
statistical systems when first (second) derivative of its Gibbs
free energy with respect to the temperature has singularity at
critical temperature $T_c$. Gibbs free energy of our CRNBH is
defined by
\begin{equation}G_+=M-T_+A_+-\Phi_+ e\end{equation} where $M$ is total
energy (mass), $A_+$ entropy and $\Phi_+$ is electric potential on
the exterior horizon $r_+$. Furthermore electric charge $e$ is a
constant.  Inserting (17), (18) and (19) the equation (22) become
\begin{equation} G_+(M,e)=\frac{\sqrt{M^2-e^2}}{2}\end{equation}
which with $e=0$ reduces to the Helmholtz free energy of a
Schwarzschild black hole as $\frac{M}{2}.$ In case of Lukewarm
type of CRNBH where $M=|e|$ the Gibbs free energy vanishes and so
the Lukewarm black hole will be has an equilibrium state
thermodynamically. Applying (18) and (23) one can obtain
\begin{equation} \frac{\partial G_+}{\partial T_+}=\frac{8\pi M(M+\sqrt{M^2-e^2})^2}{M-2\sqrt{M^2-e^2}}\end{equation} which has singularity at critical point
$\frac{M}{|e|}=\frac{2}{\sqrt{3}}.$ Its singular point is same as
 one which is obtained for heat capacity (20).
  Thus CRNBH metric exhibits with first order phase transition. Second
order derivative of the Gibbs energy (23) is obtained with respect
to temperature $T$ as
\begin{equation}\frac{\partial^2G_+}{\partial
T^2_+}=128\pi^2(M+\sqrt{M^2-e^2})^4$$$$\times\frac{(M^2+e^2-2\sqrt{M^2-e^2})}{(M-2\sqrt{M^2-e^2})^3}\end{equation}
which is not well behaved for the critical point
$M_c=\frac{2|e|}{\sqrt{3}}$. This is necessary and sufficient
conditions for the system to perform a first order phase
transition (see page 83 in ref. \citep{Rei80}). In other words the
entropy and temperature are single valued at the critical point
$M_c=\frac{2|e|}{\sqrt{3}}$ but not the heat capacity or first
order derivative of Gibbs energy with respect to temperature. Our
results agree with the work presented by
 \citep{Dav77}. In case
$1<\frac{M}{|e|}<\frac{2}{\sqrt{3}}$ the heat capacity (20)  takes
positive values which means the RN black hole is in equilibrium
with its surrounding
 heat bath, but in case $\frac{M}{|e|}>\frac{2}{\sqrt{3}}=1.1547$ it become disequilibrium because
 the heat capacity (20)  takes negative values and so a phase transition is happened there (see solid line in figure 1).
  In the latter case the CRNBH will be
 radiate its energy and is shrunk quantum mechanically by interacting the thermal Hawking radiation (see figures 9 and 10).
 So we  must
 be consider backreaction corrections of the thermal Hawking radiation
 on event horizons of  evaporating RN black hole to obtain corresponding thermodynamical quantities.
  We will do this in the sections 5 and 6. We complete now the present subsection by
recalling other claim for  phase transition of CRNBH presented by
 Meitei et al \citep{Mei10}:\\
The electric potential (19)  has not
  singular point and so dose not exhibit discontinuity
(see also Eq. (2.3) in ref.
 \citep{Lou97}), but Meitei et al are shown in ref. \citep{Mei10},
 that the CRNBH electric potential
 (19) can be rewritten as \begin{equation}
 \Phi=\bigg(\frac{\partial M}{\partial
 e}\bigg)_{T}=\frac{er_-}{2e^2-Mr_+}.\end{equation}
 This form of electric potential exhibits with discontinuity
 at the critical point $\frac{M}{|e|}=\frac{2}{\sqrt{3}}.$ Hence they claimed that the discontinuity is not physical and so
 the phase transition is apparent. The latter statement can not be correct in my opinion because of
 singularity which is happened on  the first derivative of the corresponding Gibbs free energy.
Eliminating $M$ between (17), (18) and (23) one can obtain
temperature dependent form of the Gibbs free energy and the
entropy respectively as
 follows.
  \begin{equation}T_+(G_+)=\frac{G_+}{4\pi (\sqrt{4G^2_++e^2}+2G_+)^2}\end{equation}
and
\begin{equation}A_+(G_+)=4\pi[(4G_+^2+e^2)^{\frac{1}{2}}+2G_+]^2\end{equation}
 where $G_+>0$ for all $M>|e|$ (see (23)). Their diagram are given in RHS of the
figures (3) and (4) respectively.
\section{Entropy on interior
horizon} Interior horizon $r_-=M-\sqrt{M^2-e^2}$ can not be
observed by an observer located outside of exterior horizon but
its thermodynamics properties can be take some physical statements
as follows. One can calculate entropy of interior horizon of the
CRNBH  as
\begin{equation}A_-(M,e)=4\pi
r_-^2=4\pi(M-\sqrt{M^2-e^2})^2.\end{equation}
\subsection{Negative temperatures }
Varying the entropy equation (29) with respect to $M$ and $e$ and
comparing with (16) one can obtain corresponding temperature and
electric potential on the interior horizon as follows.
\begin{equation}T_-(M,e)=-\frac{1}{8\pi}\frac{\sqrt{M^2-e^2}}{(M-\sqrt{M^2-e^2})^2}\end{equation}
and
\begin{equation}\Phi_-(M,e)=\frac{e}{M-\sqrt{M^2-e^2}}.\end{equation}
Regarding (21) we can obtain heat capacity equation for the
interior horizon as
\begin{equation}C_e^-(M,e)=-8\pi\bigg[\frac{(M-\sqrt{M^2-e^2})^2\sqrt{M^2-e^2}}{M+2\sqrt{M^2-e^2}}\bigg]
\end{equation}
which dose not exhibit with a singularity. Temperature (30) takes
negative values always and we give diagram of the solutions (29),
(30) and (32) in figure 2. Temperature, entropy and heat capacity
are plotted against $M$ with dash, dot and solid lines
respectively for particular electric charges $e=\pm1.$ Gibbs free
energy on the interior horizon is given by
\begin{equation}G_-(M,e)=M-T_-A_--\Phi_-e.\end{equation} Inserting (29), (30) and (31) explicit form of the Gibbs free energy (33) become
 \begin{equation}
 G_-(M,e)=-\frac{1}{2}\sqrt{M^2-e^2}.\end{equation} Applying (30) and (34) one obtain
\begin{equation}\frac{\partial G_-}{\partial T_-}=\frac{4\pi M(M-\sqrt{M^2-e^2})^2}{M+2\sqrt{M^2-e^2}}\end{equation}
which has not a singular point. Eliminating mass parameter $M$
between (30) and (34), temperature dependent function of Gibbs
free energy become
\begin{equation}T_-(G_-)=\frac{G_-}{4\pi
(\sqrt{4G_-^2+e^2}+2G_-)^2}\end{equation} where $G_-<0$ (see
(34)). RHS and LHS of the figure 3 describes the temperature
equations (27) and (36) against Gibbs free energy respectively for
particular charges $e=\pm1$. Negative temperature $T<0$ appears in
ordinary statistical systems having a finite energy maximum
$E_{max}$ where when entropy is continuously increasing
(decreasing) then energy and temperature decrease (increase). \\
In the present work initial mass of the CRNBH is its finite energy
maximum $E_{max}=M$ which decays in the presence of Hawking
radiation (see figures 9, 10 and 11) and we will consider its
effects on the thermodynamical properties of the CRNBH in the
sections 5 and 6. Abatement of the CRNBH entropy is happened on
the interior horizon by increasing its mass (see dot and dash
lines at figure 2). Inserting (34) the entropy equation (29) can
be rewritten as
 \begin{equation}A_-(G_-)=4\pi[(4G_-^2+e^2)^{\frac{1}{2}}+2G_-]^2.\end{equation}
Diagram of the equations (28) and (37) are plotted against Gibbs
free energy in RHS and LHS of the figure 4 respectively. RHS of
the figure shows rise of the entropy monotonously by increasing
positive values of Gibbs free energy on the exterior horizon where
the temperature takes positive values. LHS of the figure shows
decrease of entropy to a zero value on the interior horizon by
decreasing negative values of the Gibbs free energy to negative
infinity. Namely, when entropy takes a zero value on the interior
horizon then the matter counter part inside of the interior
horizon reaches to a Bose-Einstein condensation state with zero
momentum and minimum energy (see subsection 4.2 for more
discussion). Exterior horizon exhibits with a first order phase
transition for particular mass $M_c=\frac{2|e|}{\sqrt{3}}$ where
thermodynamical functions take the following critical values.
\begin{equation}A_-(M_c)=\frac{4\pi
e^2}{3},~~~A_+(M_c)=4\pi e^2\end{equation}
\begin{equation}G_{\pm}(M_c)=\pm\frac{|e|}{2\sqrt{3}}, \end{equation}\begin{equation}C^-_e(M_c)=-\frac{2\pi e^2}{3},~~~~
 C^+_e(M_c)\to\pm\infty\end{equation}
 \begin{equation}T_+(M_c)=\frac{1}{24\pi \sqrt{3}|e|},~~~~T_-(M_c)=-\frac{9}{24\pi \sqrt{3}|e|}\end{equation}
and
\begin{equation}\Phi_-(M_c)=\frac{3}{\sqrt{3}}\frac{e}{|e|},~~~~\Phi_+(M_c)=\frac{1}{\sqrt{3}}\frac{e}{|e|}.\end{equation}
In summary after than the phase transition is happened then the
system will be proceed to take a Bose-Einstein condensation single
state as follows.
 \subsection{Bose-Einstein condensation}
Photons (bosons with spin $s=1$)  transport point particle
electric field to other places in space time, from point of
quantum electrodynamic (QCD) view. Thus we can be assume inside of
the CRNBH is accumulated with more photons with total relativistic
energy $M$ and total zero spin $(s=0)$. Total number of boson
particles is conserved in a bosonic gas and it is suitable
statistical system which can be reach to a phase transition. In
the latter system when temperature decreases to its critical value
$T\to T_c$ all of boson particles take zero momentum and so
minimum energy. Corresponding entropy (degrees of freedom)
decreases and phase transition is happened at critical temperature
$T_c$ and for temperatures $T<T_c$ the bosoinc gas reaches from
gas state to a Bose-Einstein condensation state. Figure 2 (dot
line) shows decrease of entropy at negative temperatures by
increasing $M$ and so for CRNBH the Bose-Einstein
condensation state may to be happened at negative temperatures.\\
Negative temperatures have several physical features and we
address results of some suitable works as: (1) The Bose-Einstein
condensation phase transition can be happened at negative
temperatures \citep{Mos05}. (2) Dark matter perfect fluid with
negative pressure corresponds boson fields with negative
temperatures which can be support big rip singularity of the
expanding universe in the finite future \citep{Ped04}. (3)
Negative temperature can be create an attractively interacting
ensemble of ultra-cold bosons which are stable and could not
collapsed for arbitrary atom numbers
\citep{Bra13} (see also \citep{Car13}). \\
As an extension of our work it is useful to consider Hawking
radiation temperature of the CRNBH as dynamical backreaction
effects on the CRNBH itself and obtain corrected counter part for
entropy, temperature Gibbs free energy and heat capacity at
constant electric charge $e$ for final state of evaporating QRNBH
metric in the next section. In the context of  alternative gravity
theories in absence of torsion $f(R)$ and in presence of torsion
$f(T)$ effects black hole thermodynamics is studied also in the
literature by more authors which  can be addressed
 to for instance \citep{Cem11} and \citep{Gam15}
respectively (see also references therein).
\begin{figure}[tbp] \centering
    \includegraphics[width=7cm,height=7cm]{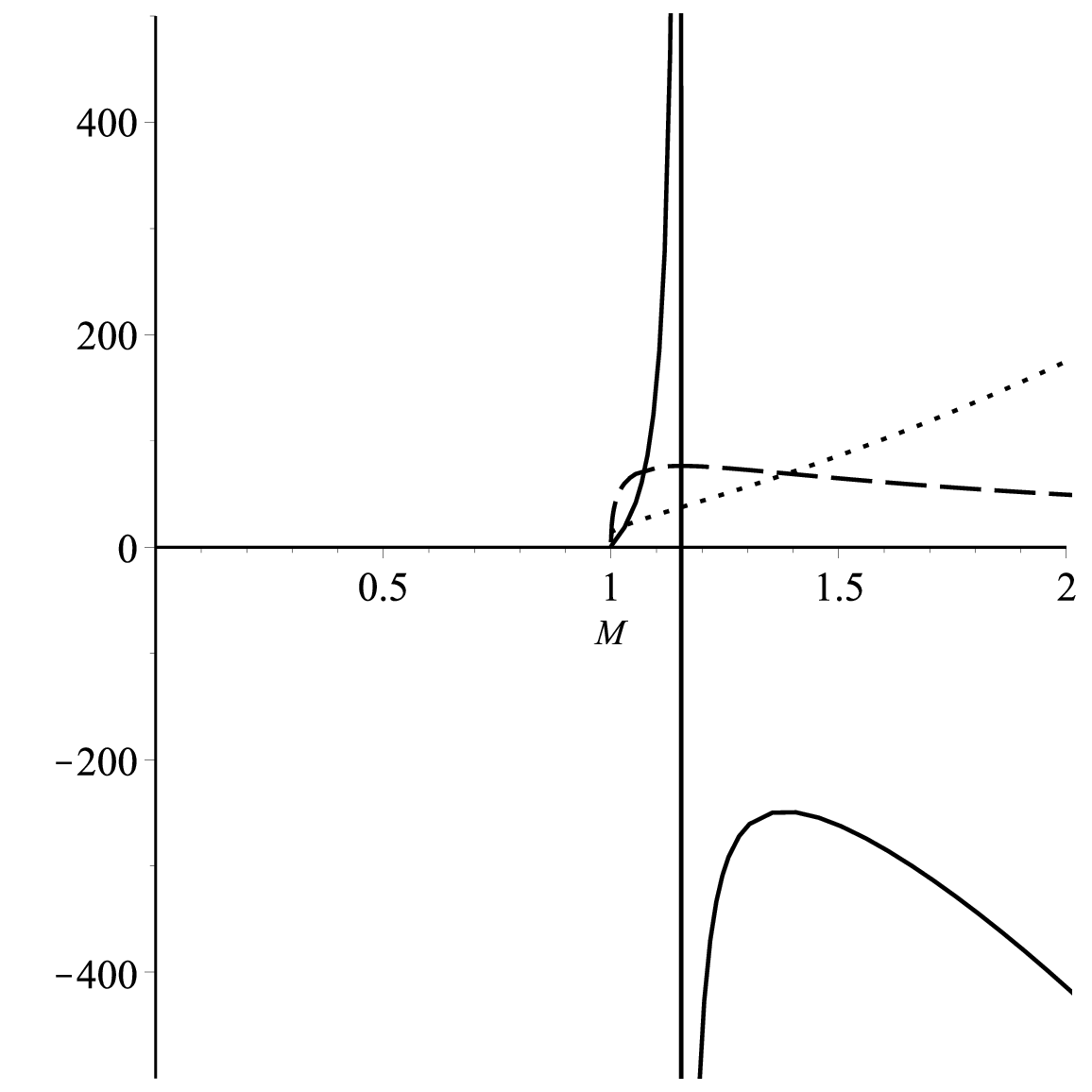}
\caption{\label{fig:epsart} Diagram of the entropy (17),
temperature (18) and heat capacity (20) are plotted against $M$
 with `dot`, `dash` and `solid`
lines respectively for particular charges $e=\pm1$. Temperature
values on the vertical axis are re-scaled as $\times10000$ but not
for entropy and heat capacity. $\Lambda-$ phase transition is
happened at positive critical temperature
$T_c(M_c)=\frac{1}{24\pi\sqrt{3}|e|}$, with
$M_c=\frac{2|e|}{\sqrt{3}}$. }
\end{figure}
\begin{figure}[tbp] \centering
    \includegraphics[width=7cm,height=7cm]{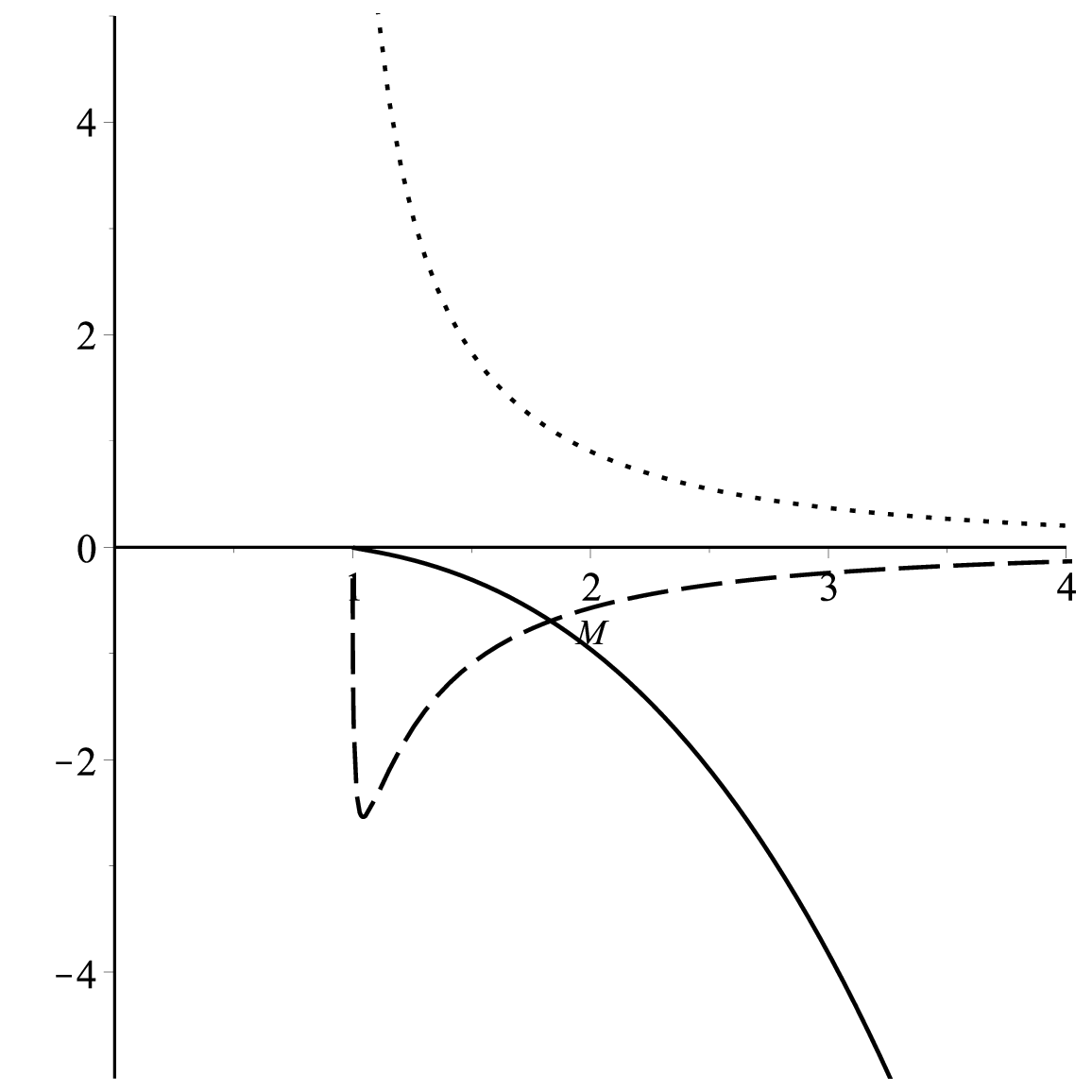}
\caption{\label{fig:epsart} Diagram of the entropy (29),
temperature (30) and heat capacity (32) are plotted against $M$
with `dot`, `dash` and `solid` lines respectively for particular
charges $e=\pm1.$ Phase transition dose not happened there but
system takes negative temperature and its entropy decreases and so
the system can be reach to Bose-Einstein condensation state.}
\end{figure}
\begin{figure}[tbp] \centering
    \includegraphics[width=7cm,height=7cm]{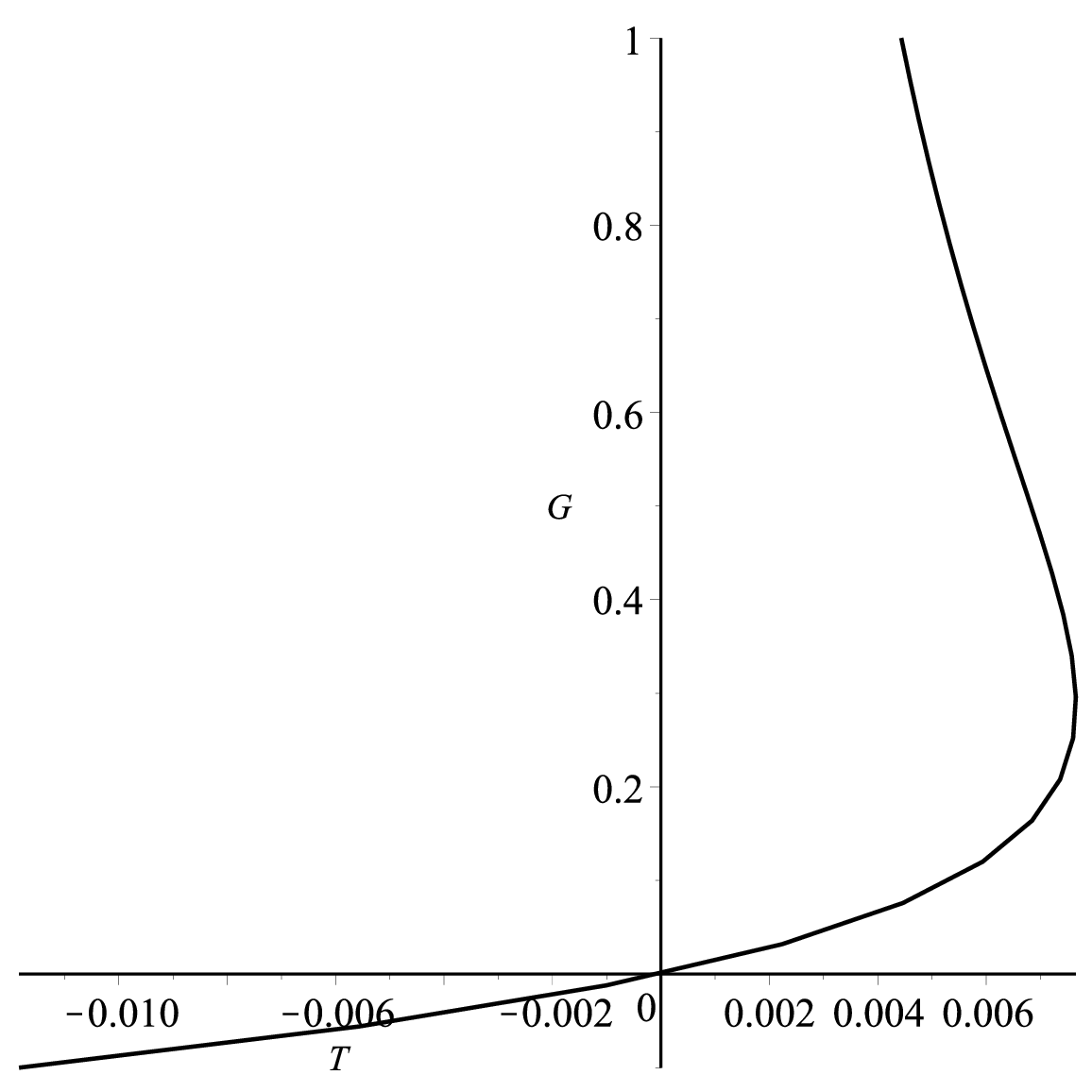}
\caption{\label{fig:epsart} LHS and RHS of the diagram describe
variation of the temperature functions (27) and (30) with respect
to negative Gibbs free energy $G_-<0$ and its positive values
$G_+>0$ respectively. In other words LHS and RHS of the diagram
denote to variation of interior and exterior horizon temperature
respectively against corresponding Gibbs free energy.}
\end{figure}
\begin{figure}[tbp] \centering
    \includegraphics[width=7cm,height=7cm]{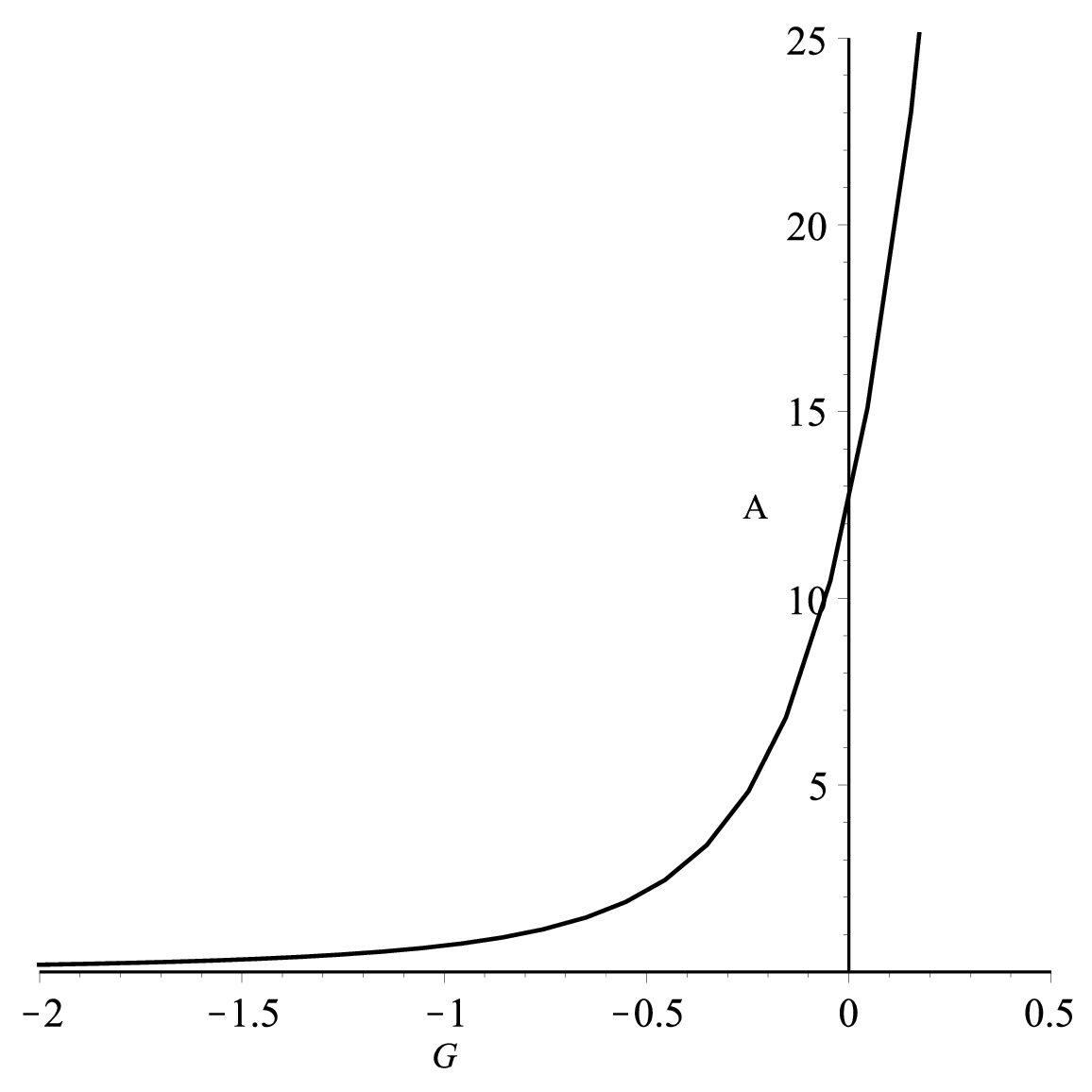}
\caption{\label{fig:epsart} Diagram of entropy functions (28) and
(37) are plotted against negative  Gibbs free energy $G_-<0$
(interior horizon) and positive Gibbs energy $G_+>0$ (exterior
horizon) respectively at LHS and RHS of the figure. }
\end{figure}
 \section{Quantum Reissner-Nordstr\"{o}m Black Hole} Using perturbation series expansion method presented by
York \citep{Yor85}, backreaction corrections of conformaly
invariant massless, charge-less quantum scalar fields is used to
solve time independent backreaction metric equation of an
evaporating QRNBH metric by Bobo et al \citep{Bob01}. They
obtained final state of the evaporating metric as remnant stable
fat black hole metric where exterior and interior quantum event
horizon radiuses $r^{QH}_{\pm}$ are obtained approximately as (see
Eqs. (68) and (69) in ref. \citep{Bob01})
\begin{equation} r^{QH}_+\approx
r_+\bigg[1+\epsilon\bigg(\frac{r_++r_-}{r_+-r_-}\bigg)
\frac{D}{M}\bigg]\end{equation} and
\begin{equation}r^{QH}_-\approx
r_-\bigg[1-\epsilon\bigg(\frac{r_++r_-}{r_+-r_-}\bigg)
\frac{D}{M}\bigg]\end{equation} where $0<\epsilon<1$ is order
parameter of perturbation series expansion of the well known
metric backreaction equation $G_{\mu\nu}=8\pi
(T^{class}_{\mu\nu}+<\hat{T}_{\mu\nu}>_{ren})$ and the cutoff
length $D$ is related to radiation energy of interacting quantum
fields as $D=E_{rad}$ (see Eq. (77) in ref.\citep{Bob01} ). In
absence of quantum field backreaction corrections $D=0$ the
perturbed event horizons radius lead to the classical values
$r_{\pm}.$ Inserting $r_{\pm}=M\pm\sqrt{M^2-e^2},$ the relations
(43) and (44) become respectively \begin{equation}
r^{QH}_{+}\simeq(M+\sqrt{M^2-e^2})\bigg(1+\frac{\epsilon
D}{\sqrt{M^2-e^2}}\bigg)\end{equation} and \begin{equation}
r^{QH}_{-}\simeq(M-\sqrt{M^2-e^2})\bigg(1-\frac{\epsilon
D}{\sqrt{M^2-e^2}}\bigg).\end{equation} Regarding
Bekeneshtein-Hawking entropy theorem (2), one can obtain exterior
and interior quantum perturbed entropy function by calculating
exterior and interior horizon surface area $A^{Q}_{\pm}(M,e)=4\pi
(r^{QH}_{\pm})^2$  as
\begin{equation}A^Q_{\pm}(M,e)\simeq4\pi (M\pm\sqrt{M^2-e^2})^2$$$$\times
\bigg(1\pm\frac{\epsilon D}{\sqrt{M^2-e^2}}\bigg)^2.\end{equation}
 Varying the above entropy functions with respect to $M$ one can obtain
temperature of exterior and interior horizons of QRNBH metric
respectively as follows.
\begin{equation}T^Q_{\pm}(M,e)\simeq\pm\frac{\sqrt{M^2-e^2}}{8\pi(M\pm\sqrt{M^2-e^2})^2}$$$$\times
\bigg\{1\pm\frac{\epsilon
D(M-2\sqrt{M^2-e^2})}{M^2-e^2}\bigg\}.\end{equation}  Using (16)
 one can write a suitable equation for the electric potential
at constant entropy  as
\begin{equation}\Phi=\bigg(\frac{\partial M}{\partial e}\bigg)_A\end{equation}
where subscript $A$ denotes to a constant entropy condition. Using
(47) at constant entropy $A^Q_{\pm}=constant,$ the equation (49)
up to second order terms $O(\epsilon^2)$ reduces to the following
form.
\begin{equation}\Phi^Q_{\pm}(M,e,D)=\bigg(\frac{\partial M}{\partial e}\bigg)_{(A,D)}=\frac{e}{M\pm\sqrt{M^2-e^2}}$$$$\times
\bigg[1+\frac{\epsilon D(\sqrt{M^2-e^2}\pm M)}{(M^2-e^2)(\mp
M^2\pm e^2-M\sqrt{M^2- e^2})}\bigg]\end{equation} where constant
Hawking radiation energy of quantum massless, charge-less scalar
field $D=E_{rad}$ is independent of $e,M$ and so its variation
with respect to $e$ and $M$ vanishes. One can calculate heat
capacity for exterior and interior horizons by applying (21)
respectively as
\begin{equation}C^{Q+}_e(M)\simeq\frac{8\pi\sqrt{M^2-e^2}(M+\sqrt{M^2-e^2})^2}{M-2\sqrt{M^2-e^2}}$$$$\times\bigg\{1
-\epsilon
D\bigg[\frac{(2M^2-3e^2+2M\sqrt{M^2-e^2})}{(M^2-e^2)(M-2\sqrt{M^2-e^2})}\bigg]\bigg\}\end{equation}
and
\begin{equation}C^{Q-}_e(M)\simeq-\frac{8\pi\sqrt{M^2-e^2}(M-\sqrt{M^2-e^2})^2}{M+2\sqrt{M^2-e^2}}$$$$\times\bigg\{1
+\epsilon
D\bigg[\frac{(3e^2-6M\sqrt{M^2-e^2})}{(M^2-e^2)(M+2\sqrt{M^2-e^2})}\bigg]\bigg\}.\end{equation}
Also one can obtain corresponding Gibbs free energy of QRNBH
metric up to second order terms $O(\epsilon^2)$ by using (22) and
(33) for the relations (47), (48) and (50) as follows.
\begin{equation}G^Q_{\pm}(M,e,D)\simeq\pm\frac{\sqrt{M^2-e^2}}{2}-\epsilon D
$$$$\times\bigg[\frac{M(M^2-e^2)(M\pm\sqrt{M^2-e^2})\mp2e^2}{2(M^2-e^2)^{\frac{3}{2}}[M\pm\sqrt{M^2-e^2}]}\bigg].\end{equation}
Inserting $\epsilon=0$ the above quantum corrected solutions lead
to the classical counterparts given in the section 4. In
perturbative approach we must be set $\epsilon D<<M$ in the
equations (43) and (44). For instance it is evaluated for unstable
circular photon orbits $r_{ph}=3M$ of Schwarzschild metric
solution in the ref. \citep{Yor85} as
\begin{equation}\epsilon\bigg(\frac{
D}{M}\bigg)\simeq3.1\times10^{-4}.\end{equation} Hence we will
consider the above sample together with $e=\pm1$ to plot diagrams
of the equations (47),
(48), (51), (52) and (53).\\
 Their  diagrams are given at figures 5, 6, 7, and 8. Comparing these figures one can result that a
 QRNBH metric will be take a Bose Einstein condensation single state at negative temperature where corresponding entropy reaches to a zero value on
 the interior horizon but phase transition is happened on the exterior horizon.
According to York`s idea \citep{Yor85} and considering time
independent regime of perturbation of quantum matter fields, our
calculations predict same thermodynamical behavior for CRNBH and
QRNBH metric to exhibit phase transition at critical temperature
and Bose-Einstein condensation state.
\begin{figure}[tbp] \centering
    \includegraphics[width=7cm,height=7cm]{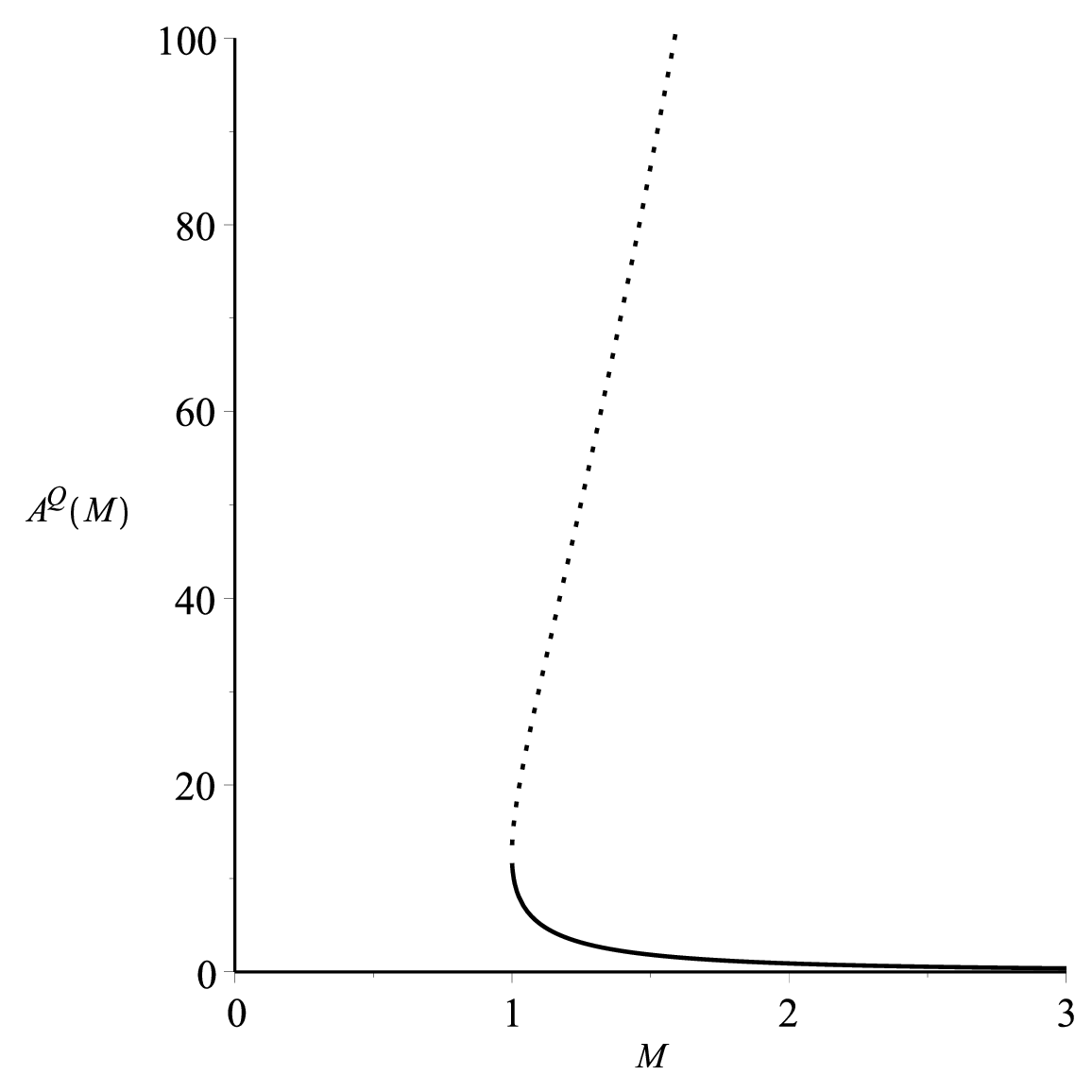}
\caption{\label{fig:epsart} Diagram of entropy $A^Q_+(M)$ and
$A^Q_-(M)$ given by (47) and defined on interior and exterior
quantum horizons are plotted against $M$ with dot and solid lines
respectively where we set $e=\pm1$ and $\epsilon
D=3.1\times10^{-4}M.$ }
\end{figure}
\begin{figure}[tbp] \centering
    \includegraphics[width=7cm,height=7cm]{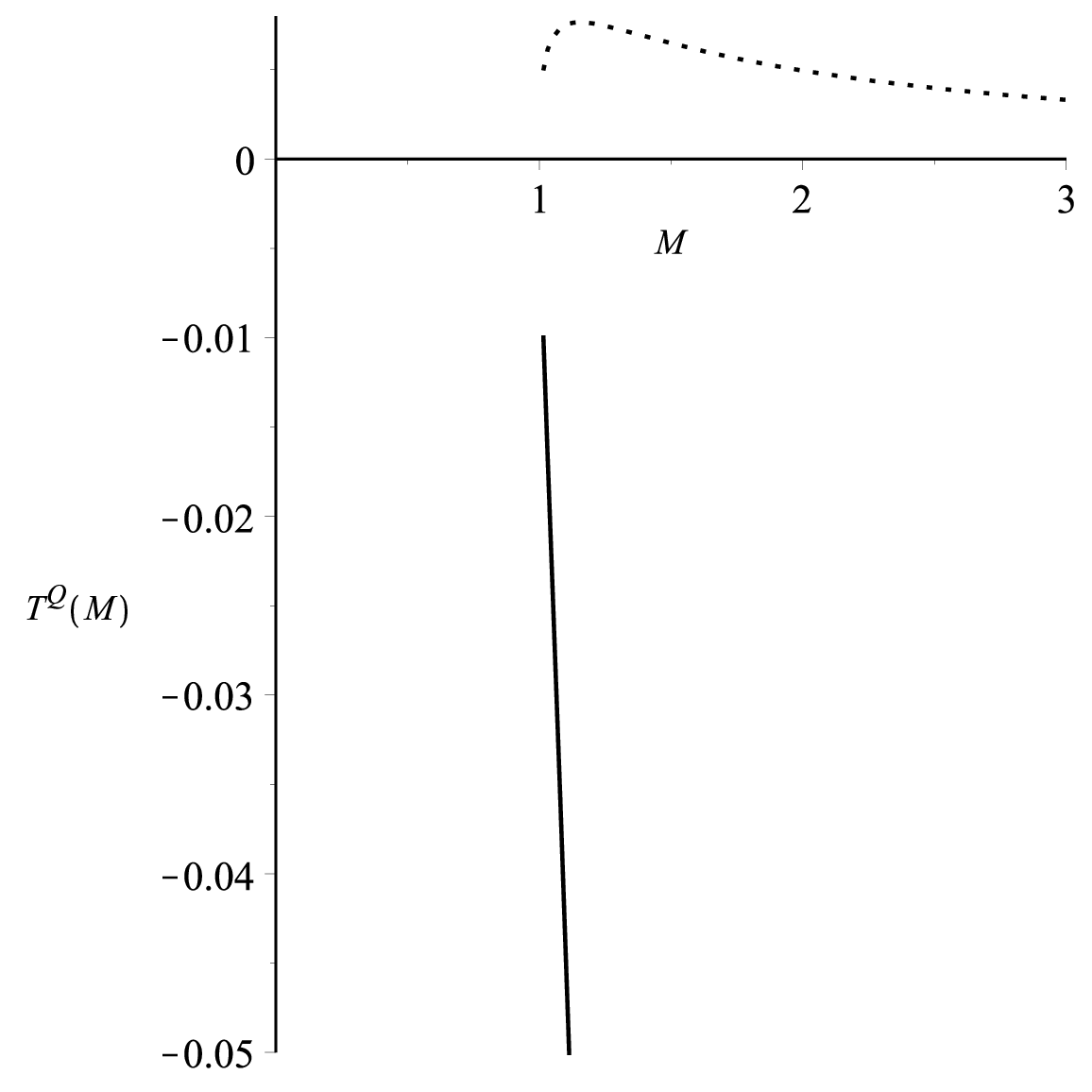}
\caption{\label{fig:epsart} Diagram of temperatures  $T^Q_+(M)$
and $T^Q_-(M)$ given by (48) and defined on interior and exterior
quantum horizons are plotted against $M$ with dot and solid lines
respectively where we set $e=\pm1$ and $\epsilon
D=3.1\times10^{-4}M.$ }
\end{figure}
\begin{figure}[tbp] \centering
    \includegraphics[width=7cm,height=7cm]{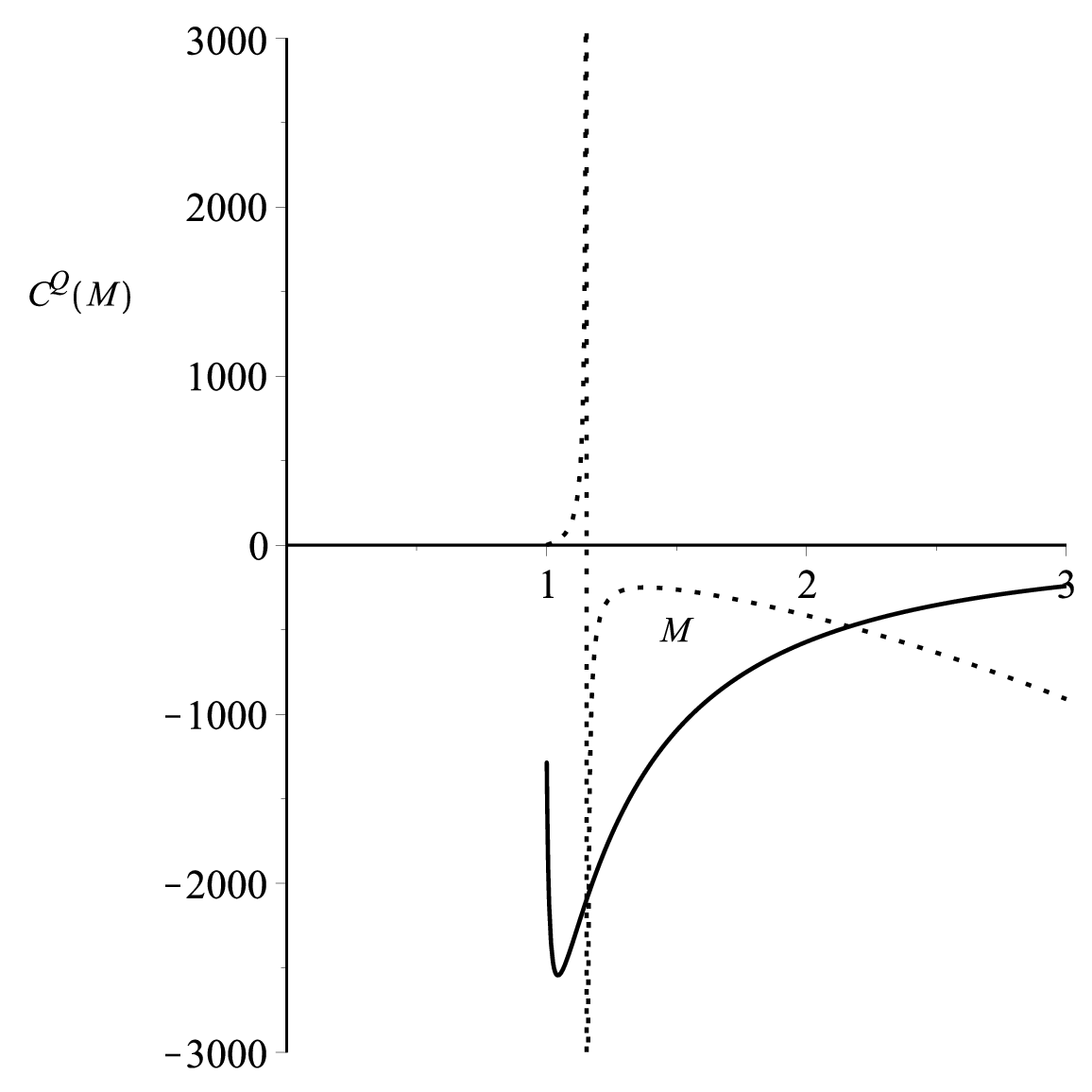}
\caption{\label{fig:epsart} Diagram of heat capacity  $C^Q_+(M)$
and $C^Q_-(M)$ given by (51) and (52) defined on interior and
exterior quantum horizons are plotted against $M$ with dot and
solid lines respectively where we set $e=\pm1$ and $\epsilon
D=3.1\times10^{-4}M$ and solid line is re-scaled as $\times1000$
with respect to dot line in the vertical axis. }
\end{figure}
\begin{figure}[tbp] \centering
    \includegraphics[width=7cm,height=7cm]{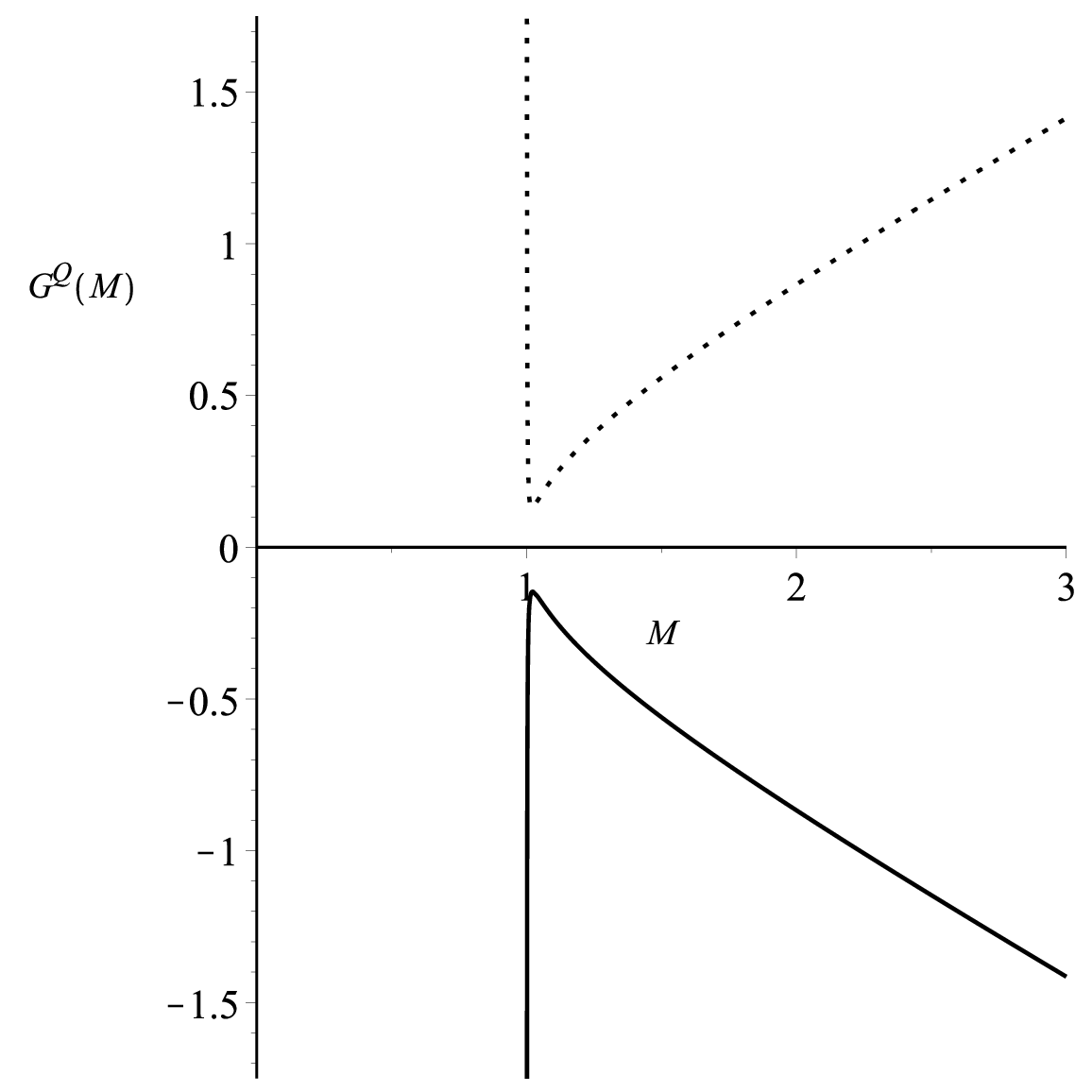}
\caption{\label{fig:epsart} Diagram of Gibbs free energy
$G^Q_+(M)$ and $G^Q_-(M)$ given by (53) and defined on interior
and exterior quantum horizons are plotted against $M$ with dot and
solid lines respectively where we set $e=\pm1$ and $\epsilon
D=3.1\times10^{-4}M.$ }
\end{figure}
\subsection{Mass loss and the switching off effect} According to all our calculations given in sections
4 and 5 for CRNBH and QRNBH metrics, the corresponding
thermodynamical variables become real quantities for
\begin{equation}M\geq|e|\end{equation} which in geometrical units is a minimum mass for evaporating RN black hole
(see all figures) because for $M<|e|$ horizons of the CRNBH and
QRNBH metrics disappear and leads to naked singularity. One can
set the above minimum mass to be equal with the Plank mass
$M_p=G^{-\frac{1}{2}}$ where semiclassical approach of quantum
gravity (quantum fields in curved space) become invalid. In the
semiclassical approach of quantum gravity Einstein tensor
(geometry) treats
 as classical field but matter stress tensor treats as quantum and so the modified Einstein metric (backreaction) equation is
 written as $G_{\mu\nu}=8\pi<\hat{T}^{quant}_{\mu\nu}
 >_{ren}.$ $<\hat{T}^{quant}_{\mu\nu}
 >_{ren}$ is renormalized (ren) expectation value of quantum matter stress tensor
 operator. Several solutions of the backreaction equation is given
 in the literature where the authors seek final state of evaporating black
 holes. They solved time independent and time dependent version of the backreaction
 equation and obtained remnant stable black holes with no naked singularities (see for instance \citep{Gha07} and references
 therein). However we calculate now QRNBH mass loss and show that it reaches to a remnant finally.\\
   Line element of evaporating RN black hole (14)
 can be written near the horizon as Vaidya form as
 \begin{equation} ds^2\simeq-\bigg(1-\frac{r_+(v)}{r}\bigg)dv^2+2drdv$$$$
 +r^2(d\theta^2+\sin^2\theta d\varphi^2)\end{equation} with the associated stress energy tensor
 \begin{equation} <\hat{T}^{quant}_{\mu\nu}>_{ren}=-\frac{1}{8\pi r^2}\frac{dr_+(v)}{dv}\delta_{\mu v}\delta_{\nu v}\end{equation}
 where $(v,r)$
 is advance
  Eddington-Finkelstein coordinates system. The black hole
  luminosity at distance $r$ is defined by \begin{equation}L(r,v)=4\pi r^2 <\hat{T}^r_v>_{ren}\end{equation}
  which by inserting (57) can be rewritten as \begin{equation}
  L=-\frac{1}{2}\frac{dr_+(v)}{dv}\end{equation}
where negative sign describes inward flux of negative energy
across the horizon which causes the black hole to shrink. The
Stefan-Boltzmann law for luminosity of black body radiation is
defined by \begin{equation}L_{SB}=A T^4\end{equation} where $A$ is
surface area of black body and $T$ is its temperature. If (59)
satisfies (60) on the apparent horizon of the QRNBH metric then we
can write mass loss equation of the black hole as follows.
\begin{equation} \frac{dr_+(v)}{dv}=-2\xi A_{+}(v)T^4_{+}(v)\end{equation}
where the normalization constant $\xi$ depends linearity on the
number of massless fields and will control the rate of
evaporation. The functions $r_+(v),$ $A_{+}(v)$ and $T_{+}(v)$ are
apparent horizon radius, its surface area and Hawking radiation
temperature respectively. They are defined explicitly by replacing
initial mass $M$ with mass function $m(v)$ for the equations (17)
and (18) such as follows. \begin{equation} A_+(v)=4\pi
r^2_+(v)\end{equation} and
\begin{equation}T_{+}(v)=\frac{\sqrt{m(v)^2-e^2}}{8\pi(m(v)+\sqrt{m^2(v)-e^2})^2}\end{equation}
where we defined
\begin{equation}r_+(v)=
m(v)+\sqrt{m(v)^2-e^2}.\end{equation} Inserting (62), (63) and
(64) the mass loss equation (61) reduces to the following form
\begin{equation}\frac{dm(v)}{dv}=-\frac{\xi}{2^{16}\pi^3}\frac{[m^2(v)-e^2]^{\frac{5}{2}}}{[m(v)+\sqrt{m^2(v)-e^2}]^7}\end{equation}
which has solution as
\begin{equation}-v^*(m^*)=({m^*}^2-1)^{\frac{7}{2}}\bigg[\frac{64}{13}{m^*}^6-\frac{1072}{143}{m^*}^4+\frac{1240}{429}{m^*}^2
$$$$-\frac{523}{3003}
\bigg]-\frac{64}{13}{m^*}^{13}+\frac{272}{11}{m^*}^{11}-\frac{152}{3}{m^*}^9+\frac{377}{7}{m^*}^7$$$$
-31{m^*}^5+9{m^*}^3-m^* +C\end{equation} where we defined
dimensionless time parameter $v^*$ and mass function $m^*(v^*)$ by
\begin{equation}v^*=\frac{\xi v(m)}{2^{16}\pi^3e}\end{equation} and \begin{equation}m^*(v^*)=\frac{m(v)}{e}.\end{equation} $C$ is
a suitable integral constant which should be set via initial
conditions of the evaporation. The mass loss solution (66) shows
that the time coordinate $v^*$ is a real parameter for $m^*\geq1$
and so we choose $v^*=0$ as the moment when evaporation is
complete so that
\begin{equation}m^*(0)=1.
\end{equation} Using (66) and (69) we must be set \begin{equation}C=\frac{16}{3003}.\end{equation}
Regarding (69) and (70) diagram of mass loss function (66) is
plotted against collapsing time $v^*$ in figure 9 for massive
QRNBH and in figure 10 for low mass QRNBH. Applying (64) and (65)
the QRNBH luminosity (59) can be rewritten as
\begin{equation}L^*(m^*)=\frac{({m^*}^2-1)^2}{({m^*}+\sqrt{{m^*}^2-1})^6}\end{equation}
 Diagram of luminosity
(71) is plotted against dimensionless mass function $m^*(v^*)$ in
figure 11.
\begin{figure}[tbp] \centering
    \includegraphics[width=7cm,height=7cm]{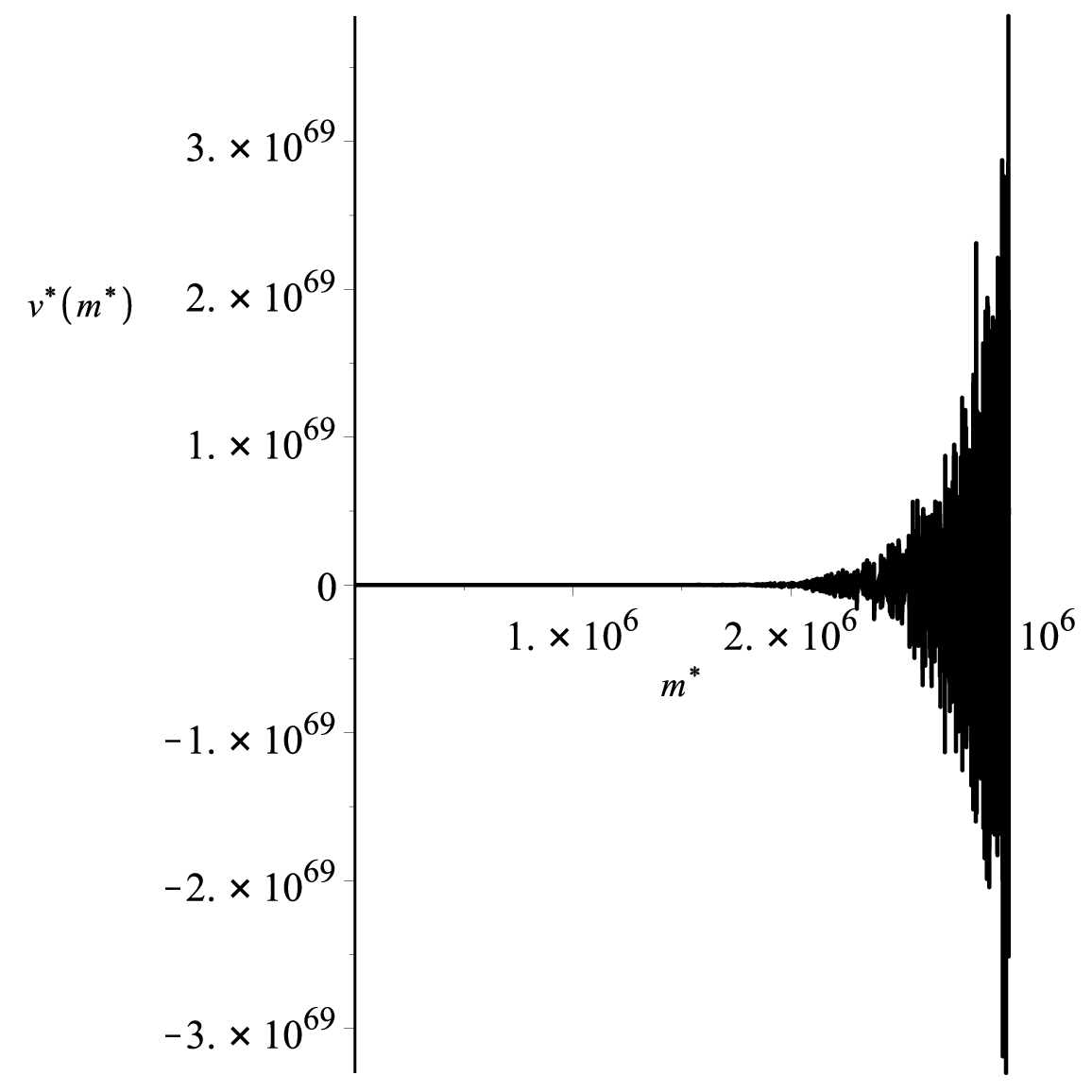}
\caption{\label{fig:epsart} Diagram of dimensionless mass function
(66) is plotted against dimensionless collapsing time $v^*$ for
massive QRNBH. }
\end{figure}
\begin{figure}[tbp] \centering
    \includegraphics[width=7cm,height=7cm]{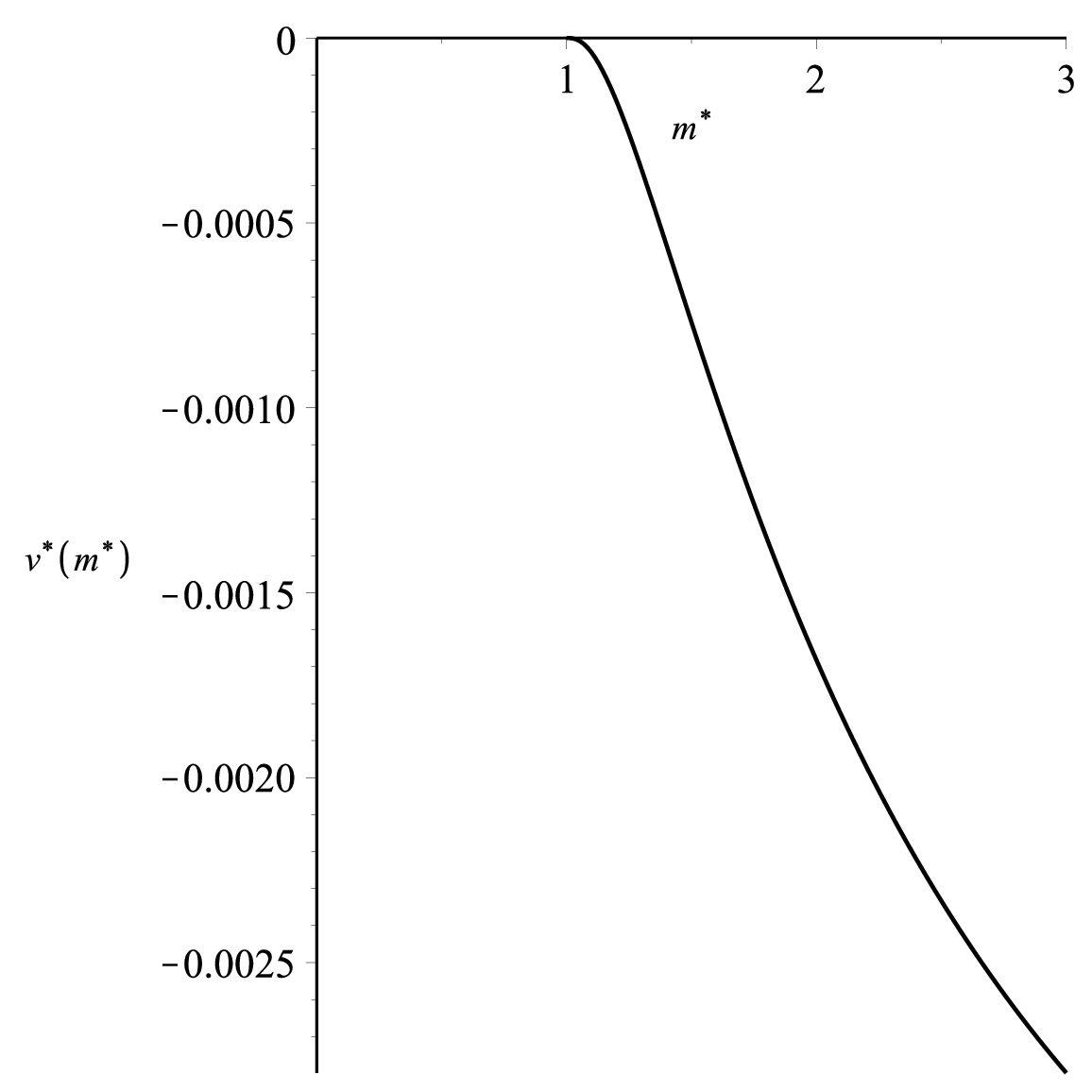}
\caption{\label{fig:epsart} Diagram of dimensionless mass function
(66) is plotted against dimensionless collapsing time $v^*$ for
low mass QRNBH. }
\end{figure}
\begin{figure}[tbp] \centering
    \includegraphics[width=7cm,height=7cm]{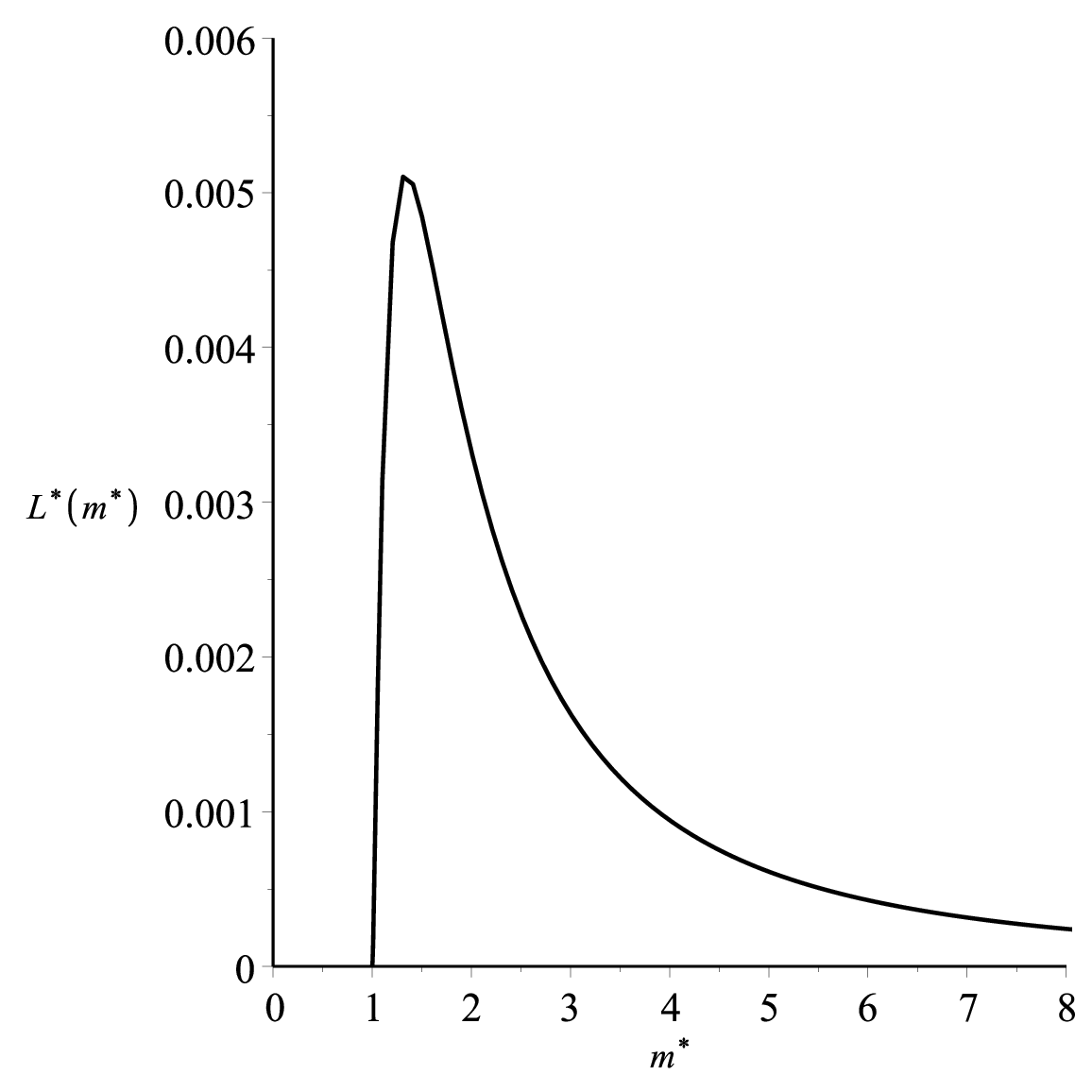}
\caption{\label{fig:epsart} Diagram of dimensionless luminosity
(71) is plotted against dimensionless mass function $m^*(v^*)$. }
\end{figure} Diagrams of the figures 9 and 10  show instability of the evaporating QRNBH which exhibits finally
to a remnant stable spherical object with non vanishing minimum
mass $m_{final}(0)=|e|$ (Lukewarm  type of RN black hole) and zero
temperature (see Eq. (63)) $T_+(m_{final})=0$ where luminosity
switches off to a zero
 value also (see figure 11). It is evident final state of QRNBH
 evaporation is conflict with final state of the Schwarzschild
 black hole evaporation reaching to a zero mass with infinite value for luminosity $L$ and
 temperature $T_+$. This is seen easily by inserting $e=0$ into the equations (63) and (65) such that
 \begin{equation}T_+(v)=\frac{1}{32\pi m(v)},~~~~e=0,\end{equation}
\begin{equation}
m(v)=\bigg[M^3-\frac{3\xi
v}{2^{23}\pi^3}\bigg]^{\frac{1}{3}}\end{equation} and
\begin{equation}L=\frac{\xi}{2^{23}\pi^3m^2(v)}\end{equation} which at the finite time
\begin{equation}v_{final}=\frac{2^{23}\pi^3 M^3}{3\xi}\end{equation}
the Schwarzschild black hole evaporation is complete with zero
final mass as
\begin{equation}m(v_{final})=0\end{equation} with
\begin{equation}(T_+,L)\to(\infty,\infty).\end{equation}
\subsection{Information loss paradox }
 Information loss paradox (ILP) is one of problems which is related to
black holes Hawking radiation. This is described as follows:\\
Usually one consider that a black hole is formed by the collapse
of matter in a pure quantum state, which then evaporates
completely into Hawking radiation. This thermal radiation
represents a transition from a pure state to a mixed state and so
violates unitarity of evolution and is forbidden in ordinary
quantum mechanics. If the Hawking radiation is assumed to be a
pure state, then would appear to require correlations between
`early` and `late` Hawking particles that have never been in
causal contact. Responding to this paradox Almheiri et al
\citep{Alm13} argue that we must be regard the following
statements: (a) the equivalence principle, (b) low energy
effective field theory or (c) the non-existence of
high-entropy `remnants` at the end of black hole evaporation.\\
Really, one should obtain a categorical answer for ILP via
un-known pure quantum gravity theory but our calculations in this
work predict an acceptable answer for the ILP at a semiclassical
approach of quantum gravity (quantum field theory in curve space).
Because we show that final state of evaporating  QRNBH reaches to
a non-vanishing mass remnant stable mini black hole with zero
temperature and zero luminosity. The latter remnant will be a
lukewarm could black hole which may be set its mass with the Plank
mass $m_p=G^{-1/2}$ where the semiclassical quantum gravity become
invalid. In the latter case there is still a tiny connection with
the macroscopic internal region. In either case, the quantum
matter field configuration in the internal region are still
correlated to external configurations. The loss of information in
the exterior part of the space time is analogous  \citep{Wal92} to
the loss of the quantum correlations which occur in any subsystem
upon tracing over the quantum state belonging to its complementary
subsystem. This loss of information is now inevitable to an
outside observer because the dynamics of the evaporation force
quantum mechanics to operate in the realm with a varying topology
(see \citep{Ren94}). But in case of quantum evaporating
Schwarzschild metric we see that the black hole mass disappear
completely and so one finds two disconnected macroscopic regions:
a quite big `baby universe` \citep{Haw88} and a Minkowskian
exterior. In summary Hawking radiation for QRNBH can be present
some suitable physical statement for ILP but not for the
evaporating quantum Schwarzschild black hole.

\section{Summary and Discussion}

In this work we studied thermodynamics of the CRNBH (QRNBH) metric
in absence (presence) of backreaction effects of Hawking particles
created by interacting of mass-less, charge-less quantum scalar
fields. We obtained that it is unstable thermodynamically and
exhibits with a first order phase transition. Evaporation reaches
to a remnant stable mini black hole called as Lukewarm black hole
where remnant mass is equal to its invariant electric charge and
temperature together with luminosity vanishes. Matter content of
evaporating QRNBH takes two different values for Gibbs free energy
against a single positive temperature on the exterior horizon but
not on the interior horizon where the corresponding temperature
takes absolutely negative values and entropy reaches to a zero
value. Thus we clime that matter content of the QRNBH contain two
different phases on the exterior horizon with rasing monotonically
entropy. But matter content located inside of interior horizon
reaches to the Bose Einstein condensation state because its
entropy reaches to a zero value at negative infinite temperature.
The first order phases transition is happened on the exterior
horizon at critical point $\frac{|e|}{M}=\frac{\sqrt{3}}{2}.$
Non-vanishing remnant mass of evaporating QRNBH can be give some
suitable physical statements for resolve the information loss
paradox. As a future work we extend aim of this work for ensemble
of CRNBH and QRNBH to seek phase transition and Bose Einstein
condensation state by regarding idea given in ref. \citep{Che07}.

\end{document}